\documentclass[a4paper,12pt]{article}
\usepackage{amssymb,amsmath}
\usepackage{bm}%
\usepackage{ascmac}
\usepackage{mathrsfs}
\usepackage{stmaryrd}
\usepackage[pdftex]{hyperref}
\hypersetup{
	colorlinks=true, 
	citecolor=blue,
	linkcolor=blue,
	urlcolor=orange,
}
\usepackage{arydshln}
\usepackage{bigdelim}
\usepackage{comment}
\usepackage{ulem}

\def\nn{\nonumber}
\def\Z2{\mathbb{Z}_2^2}
\def\bH{\mathsf{h}}
\def\bE{\mathsf{e}}
\def\DP#1#2{\hat{#1}\cdot \hat{#2}}
\def\PH#1#2{(-1)^{\hat{#1}\cdot \hat{#2}}}
\def\alg{$\Z2$-$sl_2$\ }
\def\algnospace{$\Z2$-$sl_2$}
\def\algcomma{$\Z2$-$sl_2$,\ }
\def\affine{$\mathbb{Z}_2^2$-$\widehat{sl_2}$\ }
\def\affinenospace{$\mathbb{Z}_2^2$-$\widehat{sl_2}$}
\def\affinedot{$\mathbb{Z}_2^2$-$\widehat{sl_2}$.\ }
\def\cI{\mathcal{I}}
\def\cIh{\hat{\mathcal{I}}}

\def\PBa#1#2{\{#1(y), #2(x) \}}
\def\cT{\mathcal{T}}
\def\cU{\mathcal{U}}
\def\BL{black} %chenge this to black for the final version

\title{Integrable $\Z2$-graded Extensions of the\\ Liouville and Sinh-Gordon Theories}
\author{Naruhiko Aizawa${}^{1}$\thanks{{E-mail: {\it aizawa@omu.ac.jp}}}, \quad 
Ren Ito${}^{1}$\thanks{E-mail: \textit{sd22709y@st.omu.ac.jp}}, \quad
Zhanna Kuznetsova${}^{2}$\thanks{{E-mail: {\it zhanna.kuznetsova@ufabc.edu.br}}},\\[5pt]
Toshiya Tanaka${}^{1}$\thanks{E-mail: \textit{sd23429y@st.omu.ac.jp}} \quad and\quad
Francesco Toppan${}^{3}$\thanks{{E-mail: {\it toppan@cbpf.br}}}
}
\date{\today}
\begin{document}
\thispagestyle{empty}
\maketitle
\begin{center}
{\small{	${}^{1}$\textit{Department of Physics, Graduate School of Science,
	 \\
	  Osaka Metropolitan University, Nakamozu Campus,
	  \\
	  Sakai, Osaka 599-8531 Japan.}

	  \bigskip
	  ${}^{2}$\textit{
	  UFABC, Av. dos Estados 5001, Bangu,
	  \\
	  cep 09210-580, Santo Andr\'e (SP), Brazil.
	  }

	  \bigskip
	  ${}^{3}$\textit{
	  	CBPF, Rua Dr. Xavier Sigaud 150, Urca,
	   \\
	   cep 22290-180, Rio de Janeiro (RJ), Brazil.
	  }
}}
\end{center}

%\vfill
\begin{abstract}
In this paper we present a general framework to construct integrable $\Z2$-graded extensions of classical, two-dimensional Toda and conformal affine Toda theories.  The scheme is applied to define the extended Liouville and Sinh-Gordon models; they are based on $\Z2$-graded color Lie algebras and their fields satisfy a parabosonic statististics. The  mathematical tools here introduced are the $\Z2$-graded covariant extensions of the Lax pair formalism and of the Polyakov's soldering procedure. The $\Z2$-graded Sinh-Gordon model is derived from an affine $\Z2$-graded color Lie algebra, mimicking a procedure originally introduced by Babelon-Bonora to derive the ordinary Sinh-Gordon model. The color Lie algebras under considerations are:  the $6$-generator $\Z2$-graded $sl_2$, the $\Z2$-graded  affine ${\widehat{sl_2}}$ algebra with two central extensions, the  $\Z2$-graded Virasoro algebra obtained from a Hamiltonian reduction. 
\end{abstract}

\clearpage
\setcounter{page}{1}
%%%%%%%%%%%%%%%%%%%%%%%%%%%%%%%%%%%%%%%%%%%%%%%%%%%%%%%%%%%%%%
\section{Introduction}

This paper presents a general framework to construct integrable $\Z2$-graded extensions ($\Z2 := \mathbb{Z}_2 \times \mathbb{Z}_2$) of classical, two-dimensional Toda and conformal affine Toda models. The theories under consideration possess a $\Z2$-graded color Lie algebra structure (see, for a definition, \cite{{riwy1},{riwy2}} and \cite{sch}); due to that, their component fields satisfy a parabosonic statistics. \par
Two mathematical tools are employed to prove the integrability: the first one is a  $\Z2$-graded extension
of the so-called Polyakov's soldering procedure \cite{polyak};  the second one is a $\Z2$-graded covariant extension of the Lax pair formalism introduced in  \cite{lezsav} for ordinary simple Lie algebras. \par
The above schemes are applied to construct the $\Z2$-graded versions of:\\
- the Liouville equation, derived from a $6$-generator $\Z2$-graded $sl_2$ color Lie algebra and\\
- the Sinh-Gordon model,  derived from the $\Z2$-graded  affine ${\widehat{sl_2}}$ color Lie algebra with two central extensions;  the latter case mimicks the Babelon-Bonora construction \cite{babo}  of deriving the ordinary Sinh-Gordon model
as a conformal affine Liouville theory with spontaneously broken conformal invariance.\par
Before further commenting the issues of $\Z2$-graded integrability, we briefly review the state of the art about 
the investigations regarding the \cite{{riwy1},{riwy2},{sch}} $\Z2$-graded color Lie algebras and superalgebras.
These extensions of ordinary Lie (super)algebras opened new areas of research which are of interest in both physics and mathematics. Symmetries implied by ${\mathbb Z}_2^2$-graded color Lie superalgebras appear in different physical systems, such as the de Sitter supergravity \cite{vas}, the nuclear quasi-spin \cite{jyw}, the equations of the nonrelativistic L\'evy-Leblond spinors \cite{{aktt1},{aktt2}}; further applications are the construction
of a para-Grassmann string model \cite{zhe} and the ${\mathbb Z}_2^2$-graded color extension of the super-Poincaré algebra\cite{tol2}.  \par
Color Lie (super)algebras define different types of parastatistics, see \cite{{yaji},{tol1},{stvj1},{stvj2},{top1},{top2}}. Color Lie algebras imply the presence of both bosonic and parabosonic particles, while color Lie superalgebras introduce parafermions which obey the Pauli exclusion principle. It is natural, due to the fact that color superalgebras generalize ordinary superalgebras and supersymmetry, that they have been more investigated in the literature with respect to their color Lie algebra counterparts; classical and quantum models invariant under $\Z2$-graded color Lie superalgebras  have been constructed in  \cite{{akt1},{brusigma},{brdu},{akt2},{aad},{kuto},{que}}.  The interest in the possibilities offered by physical 
theories presenting
color Lie algebra parabosons is more recent, see \cite{{kuto},{top2},{stvdjclass},{alis},{rya}}. 
More mathematical topics are the investigations of graded supergeometry (see \cite{posc,{pon}} for a review), the color
superspace formalism, see \cite{{aido1},{aizt},{aiit}}.\par
The possibility to detect paraparticles gained traction in recent years. The experimentalists learned how to simulate
paraoscillators \cite{parasim} and even engineer them in the laboratory  \cite{paraexp} by using trapped ions. On the theoretical side it has been shown, see \cite{{top1},{top2}} and also \cite{{bfrt},{top3}}, that certain results implied 
by $\Z2$-graded paraparticles  cannot be reproduced by ordinary Bose-Fermi statistics.\par
For all these reasons the field of $\Z2$-graded physics is at present quite an active area of investigations.
Obviously, the notion of integrability in the $\Z2$-graded context is one of the topics which needs to be elucidated. Some papers already started to investigate this issue, see \cite{{bruSG},{ait}}. The focus of these works has been in presenting $\Z2$-graded invariant extensions of classical, two-dimensional integrable models.
 {The present paper is the first one, as far as we know,
 	to directly address the issue of the $\Z2$-graded integrability by the mathematical tools mentioned before.
  We make some further comments about their ordinary counterparts. The Polyakov's soldering \cite{polyak} is an effective and  easy-to-implement approach (therefore, particularly suited for our scopes) to perform the Drinfeld-Sokolov \cite{drisok} Hamiltonian reduction. The Lax pair formulation allows to reconstruct, see \cite{baboto}, the solutions of the Toda field equations from chiral/antichiral free fields.  The ordinary Toda field theories defined for a simple Lie algebra $g$ are obtained as Hamiltonian reductions of the free Wess-Zumino Novikov Witten (WZNW) models whose current algebras are chiral/antichiral copies of the affine Lie algebra ${\widehat g}$.
The \cite{babo} Conformal Affine Liouville model belongs to the general class of Conformal Affine Toda theories which are obtained as Hamiltonian reduction of free WZNW models defined for ${\widehat g}$. The associated current algebras (often denoted as  ${\widehat{\widehat g}}$) are \cite{afgz}  ``double Kac-Moody algebras"; they are the classical counterparts, recovered in a limit, of the 
quantum toroidal algebras  which gained attention in recent years, see e.g. \cite{fjm}.\par
The formulation of $\Z2$-graded Toda field theories shares some common properties with the formulation of the $N=1,2$ supersymmetric Toda field theories. Just like the $N=1$ case, the simple roots are graded; this requires to introduce, for consistency, graded covariant derivatives and graded space-time coordinates (the $N=1$ superToda models are constructed, see \cite{topsupliou}, from superalgebras admitting odd simple roots). Just like the $N=2$ case, the simple roots are split into conjugate pairs. It then follows, see \cite{ivto} for the $N=2$ superToda theories, that the equations of motion are not recovered from a single Lax pair, but from two, conjugated, Lax pairs.\par
We postpone to the Conclusions further comments about the construction of $\Z2$-graded  integrable systems, the results presented in the paper and the further lines of investigation which they open. 
~\par
\par
~\par
The scheme of the paper is the following:  in Section {\bf 2} we recall the definition of $\Z2$-graded color Lie algebras and introduce a color Lie algebra extension of $sl_2$, denoted by \algcomma and its affinization \affinedot It is shown that \alg has two Casimir; a quaternionic matrix presentation of \alg which is used extensively in the paper is also given. 
In  Section {\bf 3} the $\Z2$-graded Liouville equation is derived via the Polyakov's soldering procedure. The related infinite dimensional color Lie algebras are discussed in Section {\bf 4}. It will be shown that the current algebra of the \alg case is a $\Z2$-graded version of the affine $sl_2$ algebra with a single central extension. 
As pointed out in \cite{polyak}, the Virasoro algebra is derived from the Hamiltonian reduction of $SL(2)$ gauge transformations. In the present case of \algcomma
 we obtain a $\Z2$-graded extension of the Virasoro algebra. All this is formulated in the Hamiltonian mechanics at the \textit{classical} level.  Section {\bf 5} presents the zero-curvature formulation of the $\Z2$-graded Liouville equation. Component expansions and matrix presentations of the derived equation are investigated in some detail. 
In  Section {\bf 6} the zero-curvature formulation is also applied to \affine in order to obtain the $\Z2$-graded Sinh-Gordon model. Component expansions and matrix presentations are also discussed. 
Further comments about the results of the paper, the mathematical features of the 
$\Z2$-graded integrability and further lines of investigations are given in the Conclusions.

\section{$\Z2$-graded color Lie algebras \alg and \affine}

\subsection{$\Z2$-graded color Lie algebras}

 Let us recall the definition of the $\Z2$-graded color Lie algebra. 
Let $\mathfrak{g}$ be a vector space and $\hat{a}=[a_1a_2]$ an element of $\Z2$. 
Suppose that $\mathfrak{g}$ is a direct sum of graded components 
\begin{equation}
	\mathfrak{g} = \bigoplus_{\hat{a} \in \Z2}  g_{\hat{a}}= g_{[00]} \oplus g_{[10]} \oplus g_{[01]} \oplus g_{[11]}.   
\end{equation}
If $\mathfrak{g}$ admits a bilinear operation (the graded Lie bracket), denoted by $ \llbracket \cdot, \cdot \rrbracket $ and satisfying the identities 
\begin{align}
	& \llbracket A_{\hat{a}}, B_{\hat{b}}  \rrbracket \in g_{\hat{a}+\hat{b}},
	\\
	& \llbracket A_{\hat{a}}, B_{\hat{b}} \rrbracket = -(-1)^{\hat{a}\cdot\hat{b}} \llbracket  B_{\hat{b}}, A_{\hat{a}} \rrbracket,
	\\
	& (-1)^{\hat{a}\cdot\hat{c}} \llbracket A_{\hat{a}}, \llbracket B_{\hat{b}}, C_{\hat{c}} \rrbracket \rrbracket + \PH{b}{a} \llbracket B_{\hat{b}}, \llbracket C_{\hat{c}}, A_{\hat{a}} \rrbracket \rrbracket + \PH{c}{b}\llbracket C_{\hat{c}}, \llbracket A_{\hat{a}}, B_{\hat{b}} \rrbracket \rrbracket  = 0, \label{Jacobi}
\end{align}
where $ A_{\hat{a}}, B_{\hat{a}}, C_{\hat{a}} $ are homogeneous elements of $\mathfrak{g}_{\hat{a}}$ and 
\begin{equation}
	\hat{a} + \hat{b} = [a_1+b_1 ,a_2+b_2] \in \Z2, \qquad 
	\hat{a} \cdot \hat{b} = a_1 b_2 - a_2 b_1 \in \mathbb{Z}_2,
\end{equation}
then $\mathfrak{g}$ is referred to as a $\Z2$-graded color Lie algebra. 

 It is clear from the definition that the graded Lie brackets are realized by commutators and anticommutators as follows 
 \begin{equation}
 	[\mathfrak{g}_{[00]}, \mathfrak{g}_{\hat{a}}], \qquad 
 	[\mathfrak{g}_{\hat{a}}, \mathfrak{g}_{\hat{a}}], \qquad 
 	\{ \mathfrak{g}_{\hat{a}}, \mathfrak{g}_{\hat{b}} \}, \ \hat{a} \neq \hat{b} \neq [00].
 \end{equation}

As an ordinary Lie algebra, one may define the adjoint action of $\mathfrak{g} $ on itself:
\begin{equation}
	\mathrm{ad} : \mathfrak{g} \times \mathfrak{g} \to \mathfrak{g}, \qquad 
	\mathrm{ad} A (B) := \llbracket A, B \rrbracket. 
\end{equation}
It follows from the $\Z2$-graded Jacobi identity \eqref{Jacobi} that the adjoint action is an algebraic homomorphism 
\begin{equation}
	\mathrm{ad} \llbracket A, B \rrbracket = \llbracket \mathrm{ad}A, \mathrm{ad} B \rrbracket. 
\end{equation}
Thus we have the adjoint representation of $\mathfrak{g}$ by its adjoint action.

%%%%%%%%%%%%%%%%%%%%%%%%%%%%%%%%%%%%%%%%%%%%%%%%%%%%%%%%%%%%%%%
\subsection{\algnospace : $\Z2$-graded extension of $sl_2$} \label{SEC:sl2}

In the present work we consider a $\Z2$-graded extension of $sl_2$ which is defined as follows. 
Its basis and their gradings are summarized in the table below:
\begin{equation}
	\begin{array}{c|cccc}
		& [00] & [10] & [01] & [11]
		\\[5pt] \hline
		+1 &  & E^+ & D^+ &
		\\[5pt]
		0  & H & & & Z
		\\[5pt]
		-1 &  & E^- & D^- & 
	\end{array}
\end{equation}
The first column is the eigenvalue of $ \mathrm{ad} \frac{1}{2}H. $  
The defining relations are given, in terms of (anti)commutators, by
\begin{alignat}{3}
	[H,Z] &= 0, &\qquad [H, E^{\pm}] &= \pm 2E^{\pm}, & \qquad [H, D^{\pm}] &= \pm 2 D^{\pm},
	\nn \\
	\{ Z, E^{\pm} \} &=  2 D^{\pm}, & \{Z, D^{\pm} \} &=  2E^{\pm}, & [E^+, E^-] &= H, 
	\nn \\
	\{E^{\pm}, D^{\pm}\} &= 0, & \{E^{\pm}, D^{\mp}\} &= Z, & [D^+, D^-] &= H.
	\label{sl2-def}
\end{alignat}
We also denote the basis in the ordered form
\begin{equation}
	(X^1, X^2, \dots, X^6) = (H, Z, E^+, E^-, D^+, D^-) \label{sl2-ordering}
\end{equation}
and the defining relation by
\begin{equation}
	\llbracket X^a, X^b \rrbracket = f^{ab}_{\quad c} X^c,
\end{equation}
where here and in the following the sum over the repeated indices is understood.
 
The matrices of the adjoint representation accommodate the $\Z2$-grading according to the position of the non-vanishing entries:
\begin{equation}
	\begin{array}{cccc:cc:ccc}
		& & H & Z & E^+ & E^- & D^+ & D^- & 
		\\
		H & \ldelim[{7}{*} & 00 & 11 & 10 & 10 & 01 & 01 & \rdelim]{7}{*}
		\\[2pt]
		Z & & 11 & 00 & 01 & 01 & 10 & 10 
		\\[2pt] \hdashline
		E^+ & & 10 & 01 & 00 & 00 & 11 & 11 &
		\\[2pt]
		E^- & & 10 & 01 & 00 & 00 & 11 & 11 &
		\\[2pt] \hdashline
		D^+ & & 01 & 10 & 11 & 11 & 00 & 00 & 
		\\[2pt]
		D^- & & 01 & 10 & 11 & 11 & 00 & 00 & 
	\end{array}
	\label{PosGrading}
\end{equation}
The diagonal entries are [00]-graded, so one can define the trace of the adjoint matrix. 

We now introduce two bilinear forms on \algnospace :
\begin{align}
		g^{ab} := g(X^a,X^b) := Tr(\mathrm{ad} X^a \cdot \mathrm{ad} X^b), \label{Def:Killing}
		\\
		 \eta^{ab} := \eta(X^a,X^b) := Tr(\mathrm{ad} X^a \cdot M \cdot \mathrm{ad} X^b),\label{Def:Killing11}
\end{align}
where
\begin{equation}
	M := \begin{pmatrix}
		\sigma_1 & 0 & 0 
		\\
		0 & 0 & \sigma_3
		\\
		0 & \sigma_3 & 0
	\end{pmatrix},
	\qquad 
	\llbracket \mathrm{ad} X^a, M \rrbracket = 0. 
	\label{adXM}
\end{equation}
$g^{ab}$ is the Killing form and $\eta^{ab}$ is a [11]-graded Killing form, as it can  be seen from the position of the 
non-vanishing entries of the matrices $ (g^{ab}) $ and $ (\eta^{ab}): $ 
\begin{equation}
	g = \begin{bmatrix}
		16 & \cdot & \cdot & \cdot & \cdot & \cdot 
		\\
		\cdot & 16 & \cdot & \cdot & \cdot & \cdot 
		\\
		\cdot & \cdot & \cdot & 8 & \cdot & \cdot 
		\\
		\cdot & \cdot & 8 & \cdot & \cdot & \cdot
		\\
		\cdot & \cdot & \cdot & \cdot & \cdot & 8
		\\
		\cdot & \cdot & \cdot & \cdot & 8 & \cdot
	\end{bmatrix},
	\qquad
	\eta = 
	\begin{bmatrix}
		\cdot & 16 & \cdot & \cdot & \cdot & \cdot
		\\
		16 & \cdot & \cdot & \cdot & \cdot & \cdot
		\\
		\cdot & \cdot & \cdot & \cdot & \cdot & -8
		\\
		\cdot & \cdot & \cdot & \cdot & 8 & \cdot
		\\
		\cdot & \cdot & \cdot & -8 & \cdot & \cdot
		\\
		\cdot & \cdot & 8 & \cdot & \cdot & \cdot
	\end{bmatrix}. 
	\label{BLform-matrix}
\end{equation}
The matrix $M$ is [11]-graded; it is not difficult to verify that $ M $ is the unique matrix (up to an overall constant) which $\Z2$-commute with all the adjoint matrices. 

The bilinear forms have some important properties. The first two properties are immediately seen from the definition or from \eqref{BLform-matrix} $ (\hat{a}$ denotes the grading of $X^a$):
	\begin{enumerate}
	\renewcommand{\labelenumi}{(\roman{enumi})}
	\item $ g^{ab} = 0 $ if $ \hat{a} + \hat{b} \neq [00],$ \qquad 
	$ \eta^{ab} = 0 $ if $ \hat{a} + \hat{b} \neq [11];$
	\item $ g^{ab} = g^{ba}, \qquad \eta^{ab} = (-1)^{\DP{a}{b}} \eta^{ba} ;$
	\item the non-degenerate condition
     \begin{equation}
     	g^{-1} = (g_{ab}) = \frac{1}{8} 
     	\begin{bmatrix}
     		\frac{1}{2} & \cdot & \cdot & \cdot & \cdot & \cdot
     		\\
     		\cdot & \frac{1}{2} & \cdot & \cdot & \cdot & \cdot
     		\\
     		\cdot & \cdot & \cdot & 1 & \cdot & \cdot
     		\\
     		\cdot & \cdot & 1 & \cdot & \cdot & \cdot
     		\\
     		\cdot & \cdot & \cdot & \cdot & \cdot & 1
     		\\
     		\cdot & \cdot & \cdot & \cdot & 1 & \cdot
     	\end{bmatrix}, 
     	\quad
     	\eta^{-1} = (\eta_{ab}) = \frac{1}{8} 
     	\begin{bmatrix}
     		\cdot & \frac{1}{2} & \cdot & \cdot & \cdot & \cdot
     		\\
     		\frac{1}{2} & \cdot & \cdot & \cdot & \cdot & \cdot
     		\\
     		\cdot & \cdot & \cdot & \cdot & \cdot & 1
     		\\
     		\cdot & \cdot & \cdot & \cdot & -1 & \cdot
     		\\
     		\cdot & \cdot & \cdot & 1 & \cdot & \cdot
     		\\
     		\cdot & \cdot & -1 & \cdot & \cdot & \cdot
     	\end{bmatrix};
     \end{equation}
	\item $ g(\llbracket X^a, X^b \rrbracket, X^c) = g(X^a, \llbracket X^b, X^c \rrbracket)$
	\item $ \eta(\llbracket X^a, X^b \rrbracket, X^c) = (-1)^{(\hat{a}+\hat{c})\cdot \hat{b}}\eta(X^a, \llbracket X^b, X^c \rrbracket);$
\end{enumerate}
The properties (iv) and (v)  are direct consequences of the facts that the trace is cyclic and $ \hat{a}+\hat{b} + \hat{c} = [00] $ for (iv), while  it is given by [11] for (v). 

In terms of the structure constants, (iv) and (v) are written as
\begin{align}
	f^{ab}_{\quad d} g^{dc} = f^{bc}_{\quad d} g^{ad},
	\quad
	f^{ab}_{\quad d} \eta^{dc} = (-1)^{(\hat{a}+\hat{c})\cdot \hat{b}} f^{bc}_{\quad d} \eta^{ad}.
\end{align}
They are equivalent to 
\begin{equation}
	f^{ad}_{\quad b}g_{dc} = g_{bd} f^{da}_{\quad c}, \qquad 
	f^{ad}_{\quad b} \eta_{dc} + \PH{a}{b} f^{ad}_{\quad c}\eta_{bd} = 0. 
	\label{gf-rel}
\end{equation}
Using \eqref{gf-rel}, one may verify that there exist two second order (graded) Casimir operators of \algnospace ; they are given by
\begin{align}
		C_{00} &= 8 g_{ab} X^a X^b = \frac{1}{2}(H^2+Z^2) + \{ E_+, E_-\} + \{ D_+, D_- \},
		\\
		C_{11} &= 8 \eta_{ab} X^a X^b =\frac{1}{2} \{ H, Z\} + [E^+, D^-] + [D^+,E^-],
\end{align}
where the suffices indicate their gradings. 
They have vanishing graded Lie brackets with all the elements:
\begin{equation}
	[C_{00}, X^a] = \llbracket C_{11}, X^a \rrbracket = 0, \quad \forall X^a .
	\label{CasimirComm}
\end{equation}

Finally, we introduce a matrix presentation of \algnospace which is important in the present work.  
In terms of the $2\times 2$ Identity and the Pauli matrices
\begin{equation}
	\mathbb{I}_2 = \begin{pmatrix} 
		1 & 0 \\ 0 & 1
	\end{pmatrix}, 
	\qquad
	\sigma_1 = 
	\begin{pmatrix}
		0 & 1 \\ 1 & 0
	\end{pmatrix},
	\qquad
	\sigma_2 = 
	\begin{pmatrix}
		0 & -i \\ i & 0
	\end{pmatrix},
	\qquad
	\sigma_3 = 
	\begin{pmatrix}
		1 & 0 \\ 0 & -1
	\end{pmatrix},
\end{equation}
one can introduce the $ 4 \times 4 $ complexified quaternionic matrices
\begin{align}
	    M_0 &:= \mathbb{I}_2 \otimes \mathbb{I}_2, \quad M_1 := \mathbb{I}_2 \otimes \sigma_1, \quad M_2 := \sigma_1 \otimes \sigma_2, \quad M_3 := \sigma_1 \otimes \sigma_3.
\end{align}
They satisfy for $i,j=1,2,3$ the relations (the totally antisymmetric structure constant $\epsilon_{ijk}$ is normalized so that $\epsilon_{123}=1$):
\begin{equation}
	M_i M_j = \delta_{ij} M_0+ i \epsilon_{ijk} M_k.
\end{equation}
Let $ \bH, \bE^{\pm} $ be a basis of $sl_2$ subject to the relations
\begin{equation}
	[\bH, \bE^{\pm}] = \pm 2 \bE^{\pm}, \qquad [\bE^+, \bE^-] = \bH.
\end{equation}
Then, \alg is realized by the matrices $M_k$ and $ sl_2$ as follows:
\begin{equation}
	H = M_0 \otimes \bH, \quad E^{\pm} = M_1 \otimes \bE^{\pm}, \quad 
	D^{\pm} = \pm i M_2 \otimes \bE^{\pm}, \quad Z = M_3 \otimes \bH. 
	\label{MatrixPresen}
\end{equation}
This is due to the $\Z2$-graded color Lie algebraic nature of the quaternions. 
The grading of the matrix $M_k$ is understood from \eqref{MatrixPresen}.

%%%%%%%%%%%%%%%%%%%%%%%%%%%%%%%%%%%%%%%%%%%%%%%%%%%%%%%%%%%%%%%
\subsection{\affinenospace : affine extension of $\Z2$-$sl_2$} \label{SEC:affine}

The loop extension of \alg is an infinite dimensional $\Z2$-graded color Lie algebra defined by the relations
\begin{equation}
	\llbracket X^a_n, X^b_m \rrbracket = f^{ab}_{\quad c} X^c_{n+m}, \quad n, m \in \mathbb{Z}.
\end{equation}
One may easily verify that the loop \alg algebra admits two central extensions; the first one, $c_{00}, $ is [00]-graded while the second one, $ c_{11}$, is [11]-graded:
\begin{equation}
		\llbracket X_n^a, X_m^b \rrbracket = f^{ab}_c X_{n+m}^c + \frac{n}{8} (g^{ba} c_{00} + \eta^{ba} c_{11})\, \delta_{n+m,0}. 
		\label{affineRelation}
\end{equation}
Equivalently,
\begin{alignat}{2}
	[H_n, H_m] &= 2n c_{00} \, \delta_{n+m,0}, & \qquad 
	[H_n, E_m^{\pm}] &= \pm 2 E_{n+m}^{\pm}, 
	\nonumber \\[3pt] 
	[H_n, D_m^{\pm}] &= \pm 2D_{n+m}^{\pm}, 
	&
	[H_n, Z_m] &= 2n c_{11}\,\delta_{n+m,0}, 
	\nonumber \\[3pt] 
	[E_n^{+}, E_m^-] &= H_{n+m} + n c_{00}\, \delta_{n+m,0}, &
	[E_n^{\pm}, E_m^{\pm}] &= 0, 
	\nonumber \\[3pt]
	\{E_n^{\pm}, D_m^{\pm}\}&= 0, & \{E_n^{\pm}, D_m^{\mp}\} &=  Z_{n+m}\pm n c_{11} \,\delta_{n+m,0}, 
	\nonumber \\[3pt]
	\{ E_n^{\pm}, Z_m\} &= 2 D_{n+m}^{\pm},
	&
	[D_n^{\pm}, D_m^{\pm}] &= 0, 
	\nonumber \\[3pt]
	[D_n^+, D_m^-] &= H_{n+m} + n c_{00}\, \delta_{n+m,0}, 
	&
	\{D_n^{\pm}, Z_m\} &= 2 E_{n+m}^{\pm},
	\nonumber \\[3pt]
	[Z_n, Z_m] &= 2nc_{00}\, \delta_{n+m,0} 
	\label{loopsl2-def-ext}
\end{alignat}
The central elements of a $\Z2$-graded color Lie algebra are defined as those elements having vanishing graded Lie brackets with any element. Therefore
\begin{equation}
	[ c_{00}, X^a_n] = [c_{00}, c_{11}] = \llbracket c_{11}, X^a_n \rrbracket = 0. \label{Z22center}
\end{equation}
The relation \eqref{affineRelation} is compatible with the $\Z2$-graded Jacobi identity \eqref{Jacobi}.  
Furthermore, \eqref{affineRelation} is also compatible with the graded derivations $ d_{00}, d_{11}$ defined by the relations
\begin{alignat}{2}
	[c_{\hat{00}}, d_{\hat{b}}] &=  0, 
	& \qquad 
	[d_{00}, X_n^a] &= n X_n^a,
	\nonumber \\[3pt]
	[d_{11}, H_n] &= n Z_n, &\qquad [d_{11}, Z_n] &= n H_n,
	\nonumber \\[3pt]
	\{d_{11}, E_n^{\pm}\} &= \pm n D_n, & \{ d_{11}, D^{\pm}_n\} &= \pm n E_n^{\pm}. 
	\label{Z22derivations}
\end{alignat}
We define the affine extension of \alg by
\begin{equation}
	\mathbb{Z}_2^2\text{-}\widehat{sl_2} = 
	\mathbb{C}\langle\; X_n^a \;\rangle \oplus \mathbb{C} c_{00} \oplus \mathbb{C} c_{11} \oplus \mathbb{C} d_{00} \oplus \mathbb{C} d_{11} 
\end{equation}
with the relations \eqref{affineRelation}, \eqref{Z22center} and \eqref{Z22derivations}. 

%%%%%%%%%%%%%%%%%%%%%%%%%%%%%%%%%%%%%%%%%%%%%%%%%%%%%%%%%%%%%%%%%%%%%%%%%%%%%%%%%%%%%%%%%%%%%%%%
%
\setcounter{equation}{0}
\section{$\Z2$-Liouville equation by Polyakov's soldering} \label{SEC:Soldering}

We mimick the standard procedure of soldering for deriving the $\Z2$-graded version of the Liouville equation. We introduce the $\Z2$-graded color Lie group $\Z2$-$SL(2)$ generated by the algebra \alg defined in \S\ref{SEC:sl2}. A group element of $\Z2$-$SL(2)$ is given by 
\begin{equation}
	g = \exp(\alpha_{10} E^+ + \alpha_{01} D^+) \exp(\beta_{00} H + \beta_{11} Z) \exp(\gamma_{10} E^- + \gamma_{01} D^-),
	\label{SL2elem}
\end{equation}
where the group parameters $ \alpha, \beta $ and $ \gamma$ are also $\Z2$-graded and their grading is indicated by the suffix. 
Throughout this article, the suffices 00, 10, 01, 11 indicate the $\Z2$-grading of the associated quantities.  
We assume that the group parameters are functions of the [10]-graded variables
\begin{equation}
	u, \bar{u}, \quad [u, \bar{u}] = 0. \label{10coordinates}
\end{equation}
Alternatively, one could assume that the parameters are functions of [01]-graded variables. 
It is obvious that both assumptions lead to the same equation, so we consider only the case of [10]-graded variables. 
In any case, the group parameters are regarded as graded fields on the graded coordinates.

We introduce the holomorphic and antiholomorphic WZNW-currents which are defined in terms of the group element \eqref{SL2elem}:
\begin{equation}
	J(u) := \partial_u g\cdot g^{-1}, \qquad \bar{J}(\bar u) := g^{-1} \partial_{\bar{u}} g. 
	\label{DEF:current} 
\end{equation}
By definition, the currents $J(u), \bar{J}(\bar u)$ are [10]-graded and take values in $\Z2$-$sl_2.$  
Employing the matrix presentation \eqref{MatrixPresen}, one may rearrange the components of the currents in terms of the  $sl_2$ generators $\bH, \bE^{\pm}$.  
First, the group element \eqref{SL2elem} is given by
\begin{equation}
		g = \exp( a \otimes \bE^+) \exp(b \otimes \bH) \exp(c\otimes  \bE^-),
\end{equation}
where the non-graded matrix valued fields $ a, b $ and $ c $ are defined by
\begin{align}
	\alpha_{10} E^+ + \alpha_{01} D^+ &= ( \alpha_{10} M_1 +i \alpha_{01}M_2) \otimes  \bE^+ = a \otimes  \bE^+,
	\nonumber \\
	\beta_{00} H + \beta_{11} Z &= ( \beta_{00} \mathbb{I}_2 + \beta_{11} M_3) \otimes \bH = b \otimes \bH,
	\nonumber \\
	\gamma_{10} E^- + \gamma_{01} D^- &= (\gamma_{10} M_1 -i \gamma_{01} M_2 ) \otimes \bE^- = c\otimes \bE^-.
	\label{DEF:abc}
\end{align}
It follows immediately from the definition \eqref{DEF:current} that
	\begin{align}
	J(u) &= J_+ \otimes \bE^+ + J_0 \otimes \bH + J_- \otimes \bE^-,
	\notag \\[3pt]
	\bar{J}(u) &= \bar{J}_+ \otimes \bE^+ + \bar{J}_0 \otimes \bH + \bar{J}_- \otimes \bE^-,
\end{align}
where
	\begin{alignat}{3}
	J_+ &= a_u - 2a b_u - a^2 c_u e^{-2b},
	& \quad
	J_0 &= b_u + ac_u e^{-2b},
	& \quad
	J_- &= c_u e^{-2b},
	\notag \\
	\bar{J}_+ &= a_{\bar{u}} e^{-2b}, & 
	\bar{J}_0 &=b_{\bar{u}} + c a_{\bar{u}}  e^{-2b},
	& 
	\bar{J}_- &= c_{\bar{u}}-2cb_{\bar{u}} - a_{\bar{u}} c^2 e^{-2b},
\end{alignat} 
with $ a_u := \partial_u a,\; a_{\bar{u}} := \partial_{\bar{u}}a $ etc. 
The components $J_{\pm}, J_0$ (and their conjugates) are also [10]-graded and matrix valued. 

The transformations of $ J_{\pm}, J_0$ are induced from the left action of the group element:
\begin{align}
	g &\to g' = g_{\epsilon}g,
	\notag \\
	g_{\epsilon} &= \exp(\epsilon_+ \otimes \bE^+) \exp(\epsilon_0 \otimes  \bH) \exp(\epsilon_-  \otimes \bE^-). \label{DEF:epsilon}
\end{align}
where $ \epsilon_{\pm}(u), \epsilon_0(u)$ are [00]-graded chiral functions. 
Considering the infinitesimal transformation of $g$, one may obtain
\begin{align}
	\delta_{\epsilon}J_+&=-2\epsilon_+J_0+2\epsilon_0J_++\partial_u \epsilon_+,\nonumber \\
	\delta_{\epsilon}J_0&=\epsilon_+J_--\epsilon_-J_++\partial_u \epsilon_0, \nonumber \\
	\delta_{\epsilon}J_-&=2\epsilon_-J_0-2\epsilon_0J_-+\partial_u \epsilon_-.
	\label{J-transf}
\end{align}

According to \cite{polyak} we impose constraints on the currents. 
Taking into account the grading and the matrix nature of $ J_{\pm}, J_0$, the appropriate constraints will be
\begin{alignat}{2}
	J_0(u) &= 0, & \qquad J_-(u) &= M_1, 
	\notag \\
	\bar{J}_+(u) &= -M_1, & \bar{J}_0(u)&= 0. \label{Constr2}		
\end{alignat}
The constraints on $ J_0, J_- $ and $ \bar{J}_+ $ give
\begin{equation}
    a = -b_u M_1, \qquad a_{\bar{u}} = -M_1 e^{2b}.
\end{equation}
Eliminating $ a $ from these, one obtain
\begin{equation}
	b_{u\bar{u}} M_1 = M_1 e^{2b}.
\end{equation}
Recalling the definition of $b$ in \eqref{DEF:abc}, it follows that 
\begin{equation}
 \partial_{u\bar{u}} \beta_{00}\, M_1 + \partial_{u\bar{u}} \beta_{11}\, iM_2
  = e^{2\beta_{00}} \cosh 2\beta_{11} \cdot M_1 + e^{2\beta_{00}} \sinh 2\beta_{11} \cdot iM_2.	
\end{equation}
Thus we obtain the following system of equations:
\begin{equation}
	\partial_{u\bar{u}} \beta_{00} = e^{2\beta_{00}} \cosh 2\beta_{11},
	\qquad
	\partial_{u\bar{u}} \beta_{11} = e^{2\beta_{00}} \sinh 2\beta_{11}. 
	\label{Z22-Liouville-sol}
\end{equation}
Setting $\beta_{11} = 0, $ we recover,  although the coordinate variables $u, \bar{u}$ are [10]-graded, the Liouville equation. The $\Z2$-graded nature of the equations will be discussed in \S \ref{SEC:L-comp-eq}.

%%%%%%%%%%%%%%%%%%%%%%%%%%%%%%%%%%%%%%%%%%%%%%%%%%%%%%%%%%%%%%%%%%%%%%%%%%%%%%%%%%%%%%%%%%%%%%%%
%
\setcounter{equation}{0}
\section{Current algebras}

In this Section we consider the current algebra associated with the currents given in \eqref{DEF:current}.  
This is done in the framework of classical mechanics, i.e., making use of the Poisson brackets. 
Nevertheless, we observe the existence of a central term in the Poisson Lie algebra (an example, see \cite{Anomaly}, of a classical anomaly).

\subsection{$\Z2$-graded affine algebra} \label{SEC:CACM}

Since the current $J(u)$ is \alg valued, it is expanded as
\begin{equation}
	J(u) = \cI^+_{00}E^+ + \cI^+_{11} D^+ + \cI_{10} H + \cI_{01} Z + \cI^-_{00} E^- + \cI^-_{11} D^-,
\end{equation}
where each $\mathcal{I}$ component has a $\Z2$-grading. 
The components also carry the \alg charges which are indicated by the upper suffices (no upper suffix implies charge zero). 
By the matrix presentation \eqref{MatrixPresen}, one may find the relations between the currents $J_{\pm}, J_0$ and $\mathcal{I}$'s:
\begin{align}
	J(u) &= (\cI^+_{00} M_1 + i \cI^+_{11} M_2) \otimes \bE^+ + (\cI_{10} \mathbb{I}_4 + \cI_{01} M_3) \otimes \bH + (\cI^-_{00} M_1 - i \cI^-_{11} M_2) \otimes \bE^-,
	\notag\\
	J_{\pm} &= \cI^{\pm}_{00}\, M_1 \pm i \cI^{\pm}_{11} M_2,
    \qquad
   J_0 =  \cI_{10}\, \mathbb{I}_4 + \cI_{01} M_3. \label{J2calI}
\end{align}
Similarly, the non-graded transformation parameters $\epsilon_a $ in \eqref{DEF:epsilon} are expanded as
\begin{equation}
	\epsilon_{\pm} = \varepsilon^{\pm}_{10} M_1 \pm i \varepsilon^{\pm}_{01} M_2, 
	\qquad 
	\epsilon_0 = \varepsilon_{00} \mathbb{I}_4 + \varepsilon_{11} M_3,
\end{equation}
where $\varepsilon$'s are graded transformation parameters. 
Using these, one may read off the transformation laws of the graded currents $\mathcal{I}$'s as follows:
\begin{align}
		\delta_{\epsilon} \cI^{\pm}_{00} &= 2(\pm \varepsilon_{00} \cI^{\pm}_{00} \mp \varepsilon_{10}^{\pm} \cI_{10} + \varepsilon_{01}^{\pm} \cI_{01} + \varepsilon_{11} \cI^{\pm}_{11} ) + \partial_u \varepsilon_{10}^{\pm},
		\notag\\[4pt]
		\delta_{\epsilon} \cI^{\pm}_{11} &= 2( \varepsilon_{11} \cI_{00}^{\pm} \pm \varepsilon_{01}^{\pm} \cI_{10} - \varepsilon_{10}^{\pm} \cI_{01} \pm \varepsilon_{00} \cI_{11}^{\pm} ) + \partial_u \varepsilon_{01}^{\pm}, 
		\notag\\[4pt]
		\delta_{\epsilon} \cI_{10} &= \varepsilon_{10}^+ \cI_{00}^- - \varepsilon_{10}^- \cI_{00}^+ - \varepsilon_{01}^+ \cI_{11}^- + \varepsilon_{01}^- \cI_{11}^+ + \partial_u \varepsilon_{00},
		\notag\\[4pt]
		\delta_{\epsilon} \cI_{01} &= \varepsilon_{01}^+ \cI_{00}^- + \varepsilon_{01}^- \cI_{00}^+ - \varepsilon_{10}^+ \cI_{11}^- - \varepsilon_{10}^- \cI_{11}^+ + \partial_u \varepsilon_{11}.
		\label{cItrans}
\end{align}

All the variables appearing in \eqref{cItrans}, including the coordinate $u$, are graded. 
Thus, one may consider the matrix presentation, as in \eqref{MatrixPresen}, for this system. 
The matrix presentation of the coordinate $u$ is introduced by
\begin{equation}
	u = z M_1, \quad z \in \mathbb{C} \label{uMat}
\end{equation}
It follows that the derivative with respect to $u$ is presented by
\begin{equation}
	\partial_u = M_1 \partial_z. \label{DuMat}
\end{equation}
The current $ \cI_{10}(u)$ may be expanded in a Taylor series in $u$:
\begin{align}
	\cI_{10}(u) &= \sum_{n=0} (\cI_{2n} u^{2n} + \cI_{2n+1} u^{2n+1}),
\end{align}
where the expansion coefficients have the grading:  $[\cI_{2n}] =$ [10] and $ [\cI_{2n+1}] =$ [00]. 
The [10]-graded coefficient $\cI_{2n} $ is also expressed, via the matrix $M_1$ and the non-graded constant $I_{2n}$, as $ \cI_{2n} = I_{2n} M_1$; this leads to the expression
\begin{align}
	\cI_{10}(u) &= \sum_{n=0} (I_{2n} z^{2n} + \cI_{2n+1} z^{2n+1}  ) M_1 \equiv I_1(z) M_1 \label{cIMat},
\end{align}
where $I_1(z)$ is a non-graded complex function. 
In this way one may introduce the following matrix presentation:
\begin{alignat}{4}
	\cI_{00}^{\pm} &= I_{0}^{\pm} \mathbb{I}_4, & \qquad 
	\cI_{11}^{\pm} &= I_{3}^{\pm} M_3, & \qquad 
	\cI_{10} &= I_{1} M_1, & \qquad 
	\cI_{01} &= iI_{2} M_2,
	\notag\\
	\varepsilon_{00} &= \epsilon_0 \mathbb{I}_4, & 
	\varepsilon_{11} &= \epsilon_3 M_3, & 
	\varepsilon_{10}^{\pm} &= \epsilon_1^{\pm} M_1, & 
	\varepsilon_{01}^{\pm} &=  i \epsilon_2^{\pm} M_2,  
	\label{cIMatrix}
\end{alignat}
where all $I(z)$'s and $\epsilon(z)$'s are non-graded complex functions. 
The transformation laws of the non-graded currents follow immediately from \eqref{cItrans}:
\begin{align}
	\delta_{\epsilon} I_{0}^{\pm} &= 2(\pm \epsilon_0 I_0^{\pm} \mp \epsilon_1^{\pm} I_1 - \epsilon_2^{\pm} I_2 + \epsilon_3 I_3^{\pm}) + \partial_z \epsilon_1^{\pm},
	\notag\\
	\delta_{\epsilon} I_3^{\pm} &= 2(\epsilon_3 I_0^{\pm} \pm \epsilon_2^{\pm} I_1 + \epsilon_1^{\pm} I_2 \pm \epsilon_0 I_3^{\pm}) - \partial_z \epsilon_2^{\pm},
	\notag\\
	\delta_{\epsilon} I_1 &= \epsilon_1^+ I_0^- - \epsilon_1^- I_0^+ + \epsilon_2^+ I_3^- - \epsilon_2^- I_3^+ + \partial_z \epsilon_0,
	\notag\\
	\delta_{\epsilon} I_2 &= \epsilon_2^+ I_0^- + \epsilon_2^- I_0^+ + \epsilon_1^+ I_3^- + \epsilon_1^- I_3^+-\partial_z \epsilon_3. 
	\label{CTnongr}
\end{align}

The next step consists in finding the algebra which generates the transformations \eqref{CTnongr}. 
The formulas \eqref{CTnongr} may be replaced by the Poisson bracket:
\begin{align}
	\delta_{\epsilon} Z(x) &= \frac{1}{2\pi}\oint  dy \{K(y) , Z(x)  \},
	\notag \\
	K(y)&:=s_1 \epsilon_1^- I_0^+ + s_2 \epsilon_2^- I_3^+ + s_3 \epsilon_0 I_1 + s_4 \epsilon_3 I_2 + s_5 \epsilon_1^+ I_0^- + s_6 \epsilon_2^+ I_3^- ,
	\label{TransbyGen}
\end{align} 
where $Z$ stands for the non-graded currents and the $s_i$ constants have to be determined. 
For the complex integral we take the counterclockwise contour and use the argument of a complex number as a variable of integration. 
The form  of $ K(y) $  in \eqref{TransbyGen} was determined by the following considerations. 

It is possible to introduce $\mathbb{Z}_2$-gradings to the non-graded currents in a way that is compatible with the transformation laws \eqref{CTnongr}. There are three possible assignments of $\mathbb{Z}_2$-gradings:
\begin{equation}
	\begin{array}{c|cccccccc}
		&  I_0^{\pm} & I_1 & I_2 & I_3^{\pm} & \epsilon_0 & \epsilon_1^{\pm} & \epsilon_2^{\pm} & \epsilon_3
		\\ \hline
		\mathrm{(i)} &  0 & 0 & 1 & 1 & 0 & 0 & 1 & 1	
		\\
		\mathrm{(ii)} & 1 & 0 & 0 & 1 & 0 & 1 & 1 & 0
		\\
		\mathrm{(iii)} & 1 & 0 & 1 & 0 & 0 & 1 & 0 & 1
	\end{array}
\end{equation}
The grading (iii) is not independent as it is the sum of (i) and (ii). 
The currents also have the $sl_2$ charges:
\begin{equation}
	\begin{array}{rcl}
		+1 &: & I_0^+, \ I_3^+, \ \epsilon_1^+, \ \epsilon_2^+
		\\
		0 & : & I_1, \ \; I_2, \ \; \epsilon_0, \ \epsilon_3
		\\
		-1 &: & I_0^-, \ I_3^-, \ \epsilon_1^-, \ \epsilon_2^-        
	\end{array}
\end{equation}
Their scaling dimension is $1.$ 
$ K(y) $ should have [0]-grading, zero $sl_2$ charge and scaling dimension one. 
Therefore, if we employ the $\mathbb{Z}_2$-grading (i), then \eqref{TransbyGen} is the only possible form of $K$. 
If we employ the grading (ii) and repeat the computations given below, it turns out that the results (the Poisson brackets of the currents) are the same as those derived from grading (i). 
Thus, in the following, only the grading (i) is considered. 

In order to find the Poisson brackets for the non-graded currents, by taking into account the $\mathbb{Z}_2$-grading, the $sl_2$ charges and the scaling dimension we make the following Ansatz: 
\begin{align}
	\PBa{I_0^{\pm}}{I_0^{\pm}} &= 0, 
	\notag \\ 
	\PBa{I_0^+}{I_0^-} &= a_1 I_1(y) \delta(y-x) + a_2 \partial_y \delta(y-x),
	\notag \\
	\PBa{I_0^{\pm}}{I_1} &= a_3^{\pm} I_0^{\pm}(y) \delta(y-x),
	\notag \\
	\PBa{I_0^{\pm}}{I_2} &= a_4^{\pm} I_3^{\pm}(y) \delta(y-x),
	\notag \\
	\PBa{I_0^{\pm}}{I_3^{\pm}} &= 0,
	\notag \\
	\PBa{I_0^{\pm}}{I_3^{\mp}} &= a_5^{\pm} I_2(y) \delta(y-x),
	\notag\\
	\PBa{I_1}{I_1} &= b_1 I_1(y) \delta(y-x) + b_2 \partial_y \delta(y-x),
	\notag \\
	\PBa{I_1}{I_2} &= b_3 I_2(y) \delta(y-x),
	\notag \\
	\PBa{I_1}{I_3^{\pm}} &=b_4^{\pm} I_3^{\pm}(y) \delta(y-x),
	\notag\\
	\PBa{I_2}{I_2} &= c_1 I_1(y) \delta(y-x) + c_2 \partial_y \delta(y-x),
	\notag \\
	\PBa{I_2}{I_3^{\pm}} &= c_3^{\pm} I_0^{\pm} \delta(y-x),
	\notag \\
	\PBa{I_3^{\pm}}{I_3^{\pm}} &=0,
	\notag \\
	\PBa{I_3^+}{I_3^-} &= d_1I_1(y) \delta(y-x) + d_2 \partial_y \delta(y-x),  \label{Ansatz}
\end{align}
where our convention for the delta function is
\begin{align}
	\delta(x-a) &=  \sum_{n \in \mathbb{Z}} e^{in(x-a)},
	\\
	\frac{1}{2\pi}\oint dx \delta(x) &= \frac{1}{2\pi} \int_0^{2\pi} dx \delta(x) = 1.
\end{align}
The constants $ a_i, b_i, c_i, d_i $ and $ s_i $ in \eqref{TransbyGen} are fixed by the requirement that \eqref{TransbyGen} and \eqref{Ansatz} reproduce the current transformations \eqref{CTnongr}. 
For each non-graded current, \eqref{TransbyGen} gives the following conditions:
\begin{equation}
	\begin{array}{clllll}
		Z & \multicolumn{5}{c}{\text{conditions}}
		\\ \hline
		I_0^+ & \quad 
		s_5 a_1 = 2, & s_5 a_2 = -1, & s_3 a_3^+ = -2, & s_4 a_4^+ = -2, & s_6 a_5^+ = 2
		\\[5pt]
		I_0^- & \quad 
		s_1 a_1 = 2, & s_1 a_2=-1, & s_3 a_3^- = 2, & s_4 a_4^- = -2, & s_2 a_5^- = 2
		\\[5pt]
		I_1 & \quad 
		s_1 a_3^+ = -1, &  s_5 a_3^- = 1, & s_3 b_1 = 0, &  s_3 b_2 = -1, &  s_4 b_3 = 0,
		\\[3pt]
		 & \quad 
		 s_2 b_4^+ = 1, & s_6 b_4^- = -1
		 \\[5pt]
		 I_2 & \quad
		 s_1 a_4^+ = 1, & s_5 a_4^- = 1, & s_4 c_1 =0, & s_4 c_2 = 1, & s_2 c_3^+ = -1, 
		 \\[3pt]
		 & \quad  s_6 c_3^- = -1
		 \\[5pt]
		 I_3^+ & \quad 
		 s_5 a_5^- = 2, & s_3 b_4^+ = 2, & s_4 c_3^+ = 2, &  s_6 d_1=-2, & s_6 d_2 = 1
		 \\[5pt]
		 I_3^- & \quad 
		 s_1 a_5^+ = 2, & s_3b_4^- = -2, & s_4 c_3^- = 2, & s_2 d_1 = -2, & s_2 d_2 = 1
	\end{array}
\end{equation}
Solving these conditions give the results:
\begin{alignat}{5}
	a_1 &= 2, & \quad a_2 &=-1, & \qquad a_3^{\pm} &= \mp 1, & \qquad a_4^{\pm} &=1, & \quad 
	a_5^{\pm} &= 2,
	\notag \\
	b_1 &= 0, & b_2 &= -\frac{1}{2}, & b_3 &=0, & b_4^{\pm} &= \pm 1,
	\notag \\
	c_1 &=0, & c_2 &= -\frac{1}{2}, & c_3^{\pm} &=-1, & d_1 &=-2, & d_2 &=1,
	\notag \\
	s_1 &=s_2 =1, & s_3 &=2, & s_4 &=-2, & s_5 &=s_6 =1.   
\end{alignat}
Therefore, the non-graded currents satisfy the relations
\begin{align}
	\PBa{I_0^{\pm}}{I_0^{\pm}} &= 0, 
	\notag \\ 
	\PBa{I_0^+}{I_0^-} &= 2 I_1(y) \delta(y-x) - \partial_y \delta(y-x),
	\notag \\
	\PBa{I_0^{\pm}}{I_1} &= \mp I_0^{\pm}(y) \delta(y-x),
	\notag \\
	\PBa{I_0^{\pm}}{I_2} &=  I_3^{\pm}(y) \delta(y-x),
	\notag \\
	\PBa{I_0^{\pm}}{I_3^{\pm}} &= 0,
	\notag \\
	\PBa{I_0^{\pm}}{I_3^{\mp}} &= 2 I_2(y) \delta(y-x),
	\notag\\
	\PBa{I_1}{I_1} &=  -\frac{1}{2} \partial_y \delta(y-x),
	\notag \\
	\PBa{I_1}{I_2} &= 0,
	\notag \\
	\PBa{I_1}{I_3^{\pm}} &=\pm I_3^{\pm}(y) \delta(y-x),
	\notag\\
	\PBa{I_2}{I_2} &=  -\frac{1}{2} \partial_y \delta(y-x),
	\notag \\
	\PBa{I_2}{I_3^{\pm}} &= - I_0^{\pm} \delta(y-x),
	\notag \\
	\PBa{I_3^{\pm}}{I_3^{\pm}} &=0,
	\notag \\
	\PBa{I_3^+}{I_3^-} &= -2I_1(y) \delta(y-x) +  \partial_y \delta(y-x).  
	\label{NonGrKM}
\end{align}

By expanding the currents into their modes
\begin{equation}
 I(x) = \sum_{n \in \mathbb{Z}} I_n e^{inx} \label{modeDef}
\end{equation}
we obtain the infinite dimensional Poisson-Lie algebra
\begin{alignat}{2}
	\{ I_{0,n}^{\pm},  I_{0,m}^{\pm}\} &= 0, & \qquad
	\{ I_{0,n}^+, I_{0,m}^- \} &= 2I_{1,m+n} - in \delta_{n+m,0},
	\notag \\
	\{ I_{0,n}^{\pm}, I_{1,m} \} &= \mp I_{0,n+m}^{\pm},
	&
	\{ I_{0,n}^{\pm}, I_{2,m} \} &= I_{3,n+m}^{\pm},
	\notag \\
	\{ I_{0,n}^{\pm}, I_{3,m}^{\pm} \} &= 0,
	&
	\{ I_{0,n}^{\pm}, I_{3,m}^{\mp} \} &= 2I_{2,n+m},
	\notag \\
	\{ I_{1,n}, I_{1,m}\} &= -\frac{i}{2} n \delta_{n+m,0},
	&
	\{ I_{1,n}, I_{2,m} \} &= 0,
	\notag \\
	\{ I_{1,n}, I_{3,m}^{\pm} \} &= \pm I_{3,n+m}^{\pm},
	&
	\{ I_{2,n}, I_{2,m} \} &= -\frac{i}{2} n \delta_{n+m,0},
	\notag \\
	\{  I_{2,n}, I_{3,m}^{\pm} \} &= -I_{0,n+m}^{\pm},
	&
	\{ I_{3,n}^{\pm}, I_{3,m}^{\pm} \} &= 0,
	\notag \\
	\{ I_{3,n}^+, I_{3,m}^- \} &= -2I_{1,n+m} + in \delta_{n+m,0}. 
	\label{NonGrKM2}
\end{alignat}

We now restore the $\Z2$-grading by multiplying the non-graded currents by the matrices $ M_k. $ 
However, it is impossible to restore the original $\Z2$-grading given in \eqref{cIMatrix}. 
This can be seen, for instance, from the second relation of \eqref{NonGrKM}. 
The relations \eqref{NonGrKM} or \eqref{NonGrKM2} require that
\begin{itemize}
	\item $ I_1 $ has to be assigned to the [00]-grading,
	\item $I_1^+$ and $ I_1^-$  (also $I_3^+$ and $ I_3^-$)  have to be assigned to the same grading.
\end{itemize}
It follows that the gradings of $ I_0^{\pm}, I_2, I_3^{\pm} $ should respectively be [10], [11], [01] (or their permutations). 
As an example, we assign [10], [11], [01] to $ I_0^{\pm}, I_2, I_3^{\pm}: $
\begin{equation}
		\cIh_{00,n} = I_{1,n} \mathbb{I}_4, \qquad \cIh_{11,n} = I_{2,n} M_3, \qquad 
		\cIh_{10,n}^{\pm} = I_{0,n}^{\pm} M_1, \qquad \cIh_{01,n}^{\pm} = \pm i I_{3,n}^{\pm} M_2. 
\end{equation}
Then, we obtain the following $\Z2$-graded affine Poisson-Lie algebra:
	\begin{alignat}{2}
		\{ \cIh_{00,n}, \cIh_{00,m} \} &=-\frac{i}{2}n \delta_{n+m,0}, & \qquad 
		\{ \cIh_{00,n}, \cIh_{11,m} \} &= 0,
		\notag \\
		\{ \cIh_{00,n}, \cIh_{10,m}^{\pm} \} &= \pm \cIh_{10,n+m}^{\pm}, & 
		\{ \cIh_{00,n}, \cIh_{01,m}^{\pm} \} &= \pm \cIh_{01,n+m},
		\notag \\
		\{ \cIh_{11,n}, \cIh_{11,m} \} &=-\frac{i}{2}n \delta_{n+m,0}, &
		\{ \cIh_{11,n}, \cIh_{10,m}^{\pm} \} &= \mp \cIh_{01,n+m}^{\pm},
		\notag \\
		\{ \cIh_{11,n}, \cIh_{01,m}^{\pm} \} &= \mp \cIh_{10,n+m}^{\pm}, &
		\{ \cIh_{10,n}^{\pm}, \cIh_{10,m}^{\pm} \} &= 0,
		\notag \\
		\{ \cIh_{10,n}^+, \cIh_{10,m}^- \} &= 2\cIh_{00,n+m} - in \delta_{n+m,0},
		&
		\{ \cIh_{10,n}^{\pm}, \cIh_{01,m}^{\pm} \} &= 0,
		\notag \\
		\{ \cIh_{10,n}^{\pm}, \cIh_{01,m}^{\mp} \} &= \pm 2 \cIh_{11,n+m},
		&
		\{ \cIh_{01,n}^{\pm}, \cIh_{01,m}^{\pm} \} &= 0, 
		\notag \\
		\{ \cIh_{01,n}^+, \cIh_{01,m}^- \} &= -2 \cIh_{00,n+m} + in \delta_{n+m,0}. 
		\label{RestoredAlgebra}
	\end{alignat}
The main difference between \eqref{RestoredAlgebra} and the \affine algebra introduced in \S \ref{SEC:affine} is that \eqref{RestoredAlgebra} does not have the [11]-graded central extension.

%%%%%%%%%%%%%%%%%%%%%%%%%%%%%%%%%%%%%%%%%%%%%%%%%%%%%%%%%%
\subsection{$\Z2$-graded Virasoro algebra}

We impose the constraints \eqref{Constr2} on the currents in \eqref{J2calI} and repeat the same analysis as \S \ref{SEC:CACM} to get a $\Z2$-graded extension of the Virasoro algebra. 
The constraints on the currents are given by  
\begin{align}\label{currentconstr}
	J_- = \cI^{-}_{00}\, M_1 - i \cI^{-}_{11} M_2 = M_1,
	\qquad
	J_0 =  \cI_{10}\, \mathbb{I}_4 + \cI_{01} M_3 = 0. 
\end{align}
This implies that
\begin{equation}
	\cI_{00}^- = \mathbb{I}_4, \qquad \cI_{11}^- = \cI_{10} = \cI_{01} = 0.
\end{equation}
Then, one may obtain the following transformation laws from \eqref{cItrans}:
	\begin{align}
		\delta_{\epsilon} \cI_{00}^- &= -2\varepsilon_{00} + \partial_u \varepsilon_{10}^- = 0, 
		\notag\\
		\delta_{\epsilon} \cI_{11}^- &= 2\varepsilon_{11} + \partial_u \varepsilon_{01}^- = 0,
		\notag\\
		\delta_{\epsilon} \cI_{10} &= \varepsilon_{10}^+ -\varepsilon_{10}^- \cT + \varepsilon_{01}^-\, \cU + \partial_u \varepsilon_{00} = 0,
		\notag\\
		\delta_{\epsilon} \cI_{01} &= \varepsilon_{01}^+ + \varepsilon_{01}^- \cT - \varepsilon_{10}^-\, \cU + \partial_u \varepsilon_{11} = 0
		\label{HRed}
	\end{align}
and
	\begin{align}
		\delta_{\epsilon} \cT &= 2(\varepsilon_{00} \cT + \varepsilon_{11}\, \cU) + \partial_u \varepsilon_{10}^+,
		\notag\\
		\delta_{\epsilon}\, \cU &= 2(\varepsilon_{11} \cT + \varepsilon_{00}\, \cU) + \partial_u \varepsilon_{01}^+,
		\label{delU}
	\end{align}
where we set $ \cT:= \cI_{00}^+, \; \cU := \cI_{11}^+. $

The first two relations in \eqref{HRed} give
\begin{equation}
	\varepsilon_{00} = \frac{1}{2} \partial_u \varepsilon_{10}^-, \qquad 
	\varepsilon_{11} = -\frac{1}{2} \partial_u \varepsilon_{01}^-.
\end{equation}
Substituting these expressions into the third and fourth relations in \eqref{HRed} one obtains
\begin{align}
	\varepsilon_{10}^+ &= \varepsilon_{10}^- \cT - \varepsilon_{01}^- \cU - \frac{1}{2} \partial_u^2 \varepsilon_{10}^-,
	\notag \\
	\varepsilon_{01}^+ &= -\varepsilon_{01}^- \cT + \varepsilon_{10}^- \cU + \frac{1}{2} \partial_u^2 \varepsilon_{01}^-.
\end{align}
With these relations we have the transformation laws of $\cT $ and $\cU$:
	\begin{align}
		\delta_{\epsilon} \cT &= 2(\partial_u \varepsilon_{10}^-) \cT + \varepsilon_{10}^-\partial_u \cT - \frac{1}{2} \partial_u^3 \varepsilon_{10}^- - 2(\partial_u \varepsilon_{01}^-) \cU + \varepsilon_{01}^- \partial_u \cU,
		\notag \\
		\delta_{\epsilon}\, \cU &= 2(\partial_u \varepsilon_{10}^-) \cU + \varepsilon_{10}^- \partial_u \cU + \frac{1}{2} \partial_u^3 \varepsilon_{01}^- - 2(\partial_u \varepsilon_{01}^-) \cT + \varepsilon_{01}^- \partial_u \cT.
		\label{Vir-transf}
	\end{align}
We now introduce the matrix presentation
\begin{equation}
	u = z M_1,  \qquad \cT = T\, \mathbb{I}_4,  \qquad \cU = U M_3, \qquad \varepsilon_{10}^- = \epsilon_1 M_1, \qquad \varepsilon_{01}^- = i\epsilon_2 M_2,
\end{equation}
where $z, T, U $ and $ \epsilon_k$ are [00]-graded.
The transformation laws of the non-graded currents are readily obtained from \eqref{Vir-transf}:
\begin{align}
	\delta_{\epsilon} T = 2(\partial_z \epsilon_1) T + \epsilon_1 \partial_z T - \frac{1}{2} \partial_z^3 \epsilon_1 + 2(\partial_z \epsilon_2) U + \epsilon_2 \partial_z U,
	\notag \\
	\delta_{\epsilon} U = 2(\partial_z \epsilon_1) U + \epsilon_1 \partial_z U - \frac{1}{2} \partial_z^3 \epsilon_2 + 2(\partial_z \epsilon_2) T + \epsilon_2 \partial_z T. 
	\label{Vir-transf2}
\end{align}
We note that the non-graded currents have a $\mathbb{Z}_2$-grading which is compatible with the relations \eqref{Vir-transf2}
\begin{equation}
	0 \;:\  T, \ \epsilon_1, \qquad \qquad 1 \;:\  U, \ \epsilon_2.
\end{equation}
Their scaling dimension is two. 
Taking into account these observations, we set the following Ansatz:
\begin{align}
	\{ T(y), T(x) \} &= a_1 T'(y) \delta(y-x) + a_2 T(y) \delta'(y-x) + a_3 \delta'''(y-x),
	\notag\\
	\{ U(y), U(x) \} &= b_1 T'(y) \delta(y-x) + b_2 T(y) \delta'(y-x) + b_3 \delta'''(y-x),
	\notag \\
	\{ T(y), U(x) \} &= c_1 U'(y) \delta(y-x) + c_2 U(y) \delta'(y-x),
\end{align}
where the prime  denotes the derivative with respect to $y.$ 
The constants  $a_i, b_i $ and $ c_i$ are determined by the equivalence of the transformation laws \eqref{Vir-transf2} and the relation 
\begin{equation}
	\delta_{\epsilon} Z(x) = \frac{1}{2\pi} \oint dy \{ \epsilon_1 T(y) + \epsilon_2 U(y), Z(x) \},
	\quad Z = T, U.
\end{equation}
This condition uniquely determines the constants:
\begin{equation}
	a_1 = b_1 = c_1 = -1, \qquad a_2 = b_2 = c_2 =-2, \qquad a_3 = b_3 = \frac{1}{2}.
\end{equation}
We obtain the Poisson-Lie algebra of the non-graded currents:
\begin{align}
	\{ T(y), T(x) \} &= -T'(y) \delta(y-x) -2 T(y) \delta'(y-x) + \frac{1}{2} \delta'''(y-x),
	\notag\\
	\{ U(y), U(x) \} &= -T'(y) \delta(y-x) -2 T(y) \delta'(y-x) + \frac{1}{2} \delta'''(y-x),
	\notag \\
	\{ T(y), U(x) \} &= -U'(y) \delta(y-x) -2 U(y) \delta'(y-x).
\end{align}
This algebra contains, as expected, a Virasoro subalgebra. 
Expanding the currents according to \eqref{modeDef}, we see that the modes satisfy the algebraic relations:
\begin{align}
	\{T_n, T_m  \} &= i(m-n) T_{n+m} + \frac{in^3}{2} \delta_{n+m,0},
	\notag \\
	\{U_n, U_m  \} &= i(m-n) T_{n+m} + \frac{in^3}{2} \delta_{n+m,0},
	\notag \\
	\{ T_n, U_m \} &= i(m-n) U_{n+m}. \label{NG-doubleVir}
\end{align}
The restoration of the $\Z2$-grading is straightforward since the only non-empty sectors have [00] and [11] grading. 
We set
\begin{equation}
	\cT_n := T_n \mathbb{I}_4, \qquad \cU_n := U_n M_3;
\end{equation}
then $ \cT_n, \cU_n$ satisfy the same relations as \eqref{NG-doubleVir}. 
{\color{\BL}{The current $\cT$ is a Virasoro field, while $\cU$ is a primary field of dimension $2$. The fact that the only surviving currents are [00] and [11]-graded could have been anticipated from the \eqref{currentconstr} constraints.  This construction, which does not produce from Hamiltonian reduction [10] and [01]-graded currents, is a consequence of $\Z2$-$sl_2$ being a color Lie algebra. In the Conclusions we make comments about the $\Z2$-graded color Lie superalgebras Hamiltonian reductions.  It should be mentioned that a $\Z2$-graded color Lie superalgebra extension of the Virasoro algebra has been discussed, in a different context, in \cite{zhe}.  }}

%%%%%%%%%%%%%%%%%%%%%%%%%%%%%%%%%%%%%%%%%%%%%%%%%%%%%%%%%%%%%%%%%%%%%%%%%%%%%%%%%%%%%%%%%%%%%%%%
%
\setcounter{equation}{0}
\section{Zero-curvature formulation of $\Z2$-Liouville equation}

\subsection{Derivation of the $\Z2$-Liouville equation}

Let us introduce the [10] and [01]-graded coordinates 
\begin{equation}
	[10] \ u, \; \bar{u}, \qquad [01]\ v, \; \bar{v}
\end{equation}
and the fields $\Phi, \Psi$ with values in the Cartan subalgebra of \algnospace :
\begin{align}
	\Phi(u,\bar{u}) &= \frac{1}{2} \big( \varphi_{00}(u,\bar{u}) H + \varphi_{11}(u,\bar{u}) Z\big),
	\\
	\Psi(v,\bar{v}) &= \frac{1}{2} \big( \psi_{00}(v,\bar{v}) H + \psi_{11}(v,\bar{v}) Z\big).
\end{align}
We assume that the fields $\Phi$ and $\Psi$ are [00]-graded; it follows that the component fields have  a non-trivial grading. 

Following the general construction of the Toda systems we define
\begin{alignat}{2}\label{laxes}
		L_u&=-\partial_u\Phi + e^{\mathrm{ad} \Phi} E^+, &\qquad 
		L_{\bar{u}} &=\partial_{\bar{u}}\Phi + e^{-\mathrm{ad} \Phi}E^-,
		\notag \\
		L_v &=-\partial_v\Psi + e^{\mathrm{ad} \Psi}D^+, &
		L_{\bar{v}}&=\partial_{\bar{v}}\Psi + e^{-\mathrm{ad}\Psi} D^-,
\end{alignat}
where $ e^{\pm \mathrm{ad} \Phi} X = e^{\pm \Phi} X e^{\mp \Phi}. $ 
We then consider the following linear system for $ T \in \Z2$-$SL(2)$:
\begin{equation}
	(\partial_u - L_u) T = 0, \qquad (\partial_{\bar{u}} - L_{\bar{u}}) T = 0. \label{LinSys}
\end{equation}
Similar relations for $ L_v, L_{\bar{v}}$ are also introduced; we do not need to write them explicitly since the following procedure is applied to them as well. 
The compatibility of the two equations in \eqref{LinSys} gives the zero-curvature condition which has the same form as in the non-graded case: 
\begin{equation}
		\partial_{\bar{u}} L_u - \partial_u L_{\bar{u}} + [ L_u, L_{\bar{u}}] = 0. \label{ZeroC-sl2}
\end{equation}
After straightforward computations one can see that  \eqref{ZeroC-sl2} is equivalent to
\begin{equation}
	2 \partial_{u\bar{u}}\Phi = e^{2\varphi_{00}} (\cosh2\varphi_{11}\cdot H + \sinh2\varphi_{11}\cdot Z),
\end{equation}
which gives the $\Z2$-Liouville equation
\begin{equation}
		\partial_{u\bar{u}} \varphi_{00} = e^{2\varphi_{00}} \cosh 2\varphi_{11}, 
		\qquad
		\partial_{u\bar{u}} \varphi_{11} = e^{2\varphi_{00}} \sinh 2\varphi_{11}.
		\label{Z22-L1}
\end{equation}
These equations are identical to the ones obtained in \S \ref{SEC:Soldering} by Polyakov's soldering.

A similar set of equations is obtained from $ L_v, L_{\bar v}:$
\begin{equation}
		\partial_{v\bar{v}} \psi_{00} = e^{2\psi_{00}} \cosh 2\psi_{11}, 
		\qquad
		\partial_{v\bar{v}} \psi_{11} = e^{2\psi_{00}} \sinh 2\psi_{11}. \label{Z22-L2}
\end{equation}

%%%%%%%%%%%%%%%%%%%%%%%%%%%%%%%%%%%%%%%%%%%
\subsection{Equations for the component fields} \label{SEC:L-comp-eq}

We examine the $\Z2$-graded nature of the equations in \eqref{Z22-L1}. 
To this end, we expand the fields $ \varphi_{00}, \varphi_{11} $ in a power series of $u, \bar{u}. $ 
Since $y:=u^2$, $\bar y:=\bar u^2$ are [00]-graded and commute with all other variables, it may be natural to rearrange the power series in a linear combination of the functions of $y, \bar{y} $ as follows:
\begin{align}
	\varphi_{00}(u, \bar{u})&=a_{00}(y,\bar y)+u a_{10}(y,\bar y) +\bar u b_{10}(y,\bar y) + u \bar u b_{00}(y,\bar y),
	\notag \\
	\varphi_{11}(u,\bar{u}) &= a_{11}(y,\bar{y}) + u a_{01}(y,\bar{y}) + \bar{u} b_{01}(y,\bar{y}) + u \bar{u} b_{11}(y,\bar{y}). \label{uubarcomp}
\end{align}
We call the functions on the right hand side \textit{component fields} of $\varphi_{00}$ (or $\varphi_{11})$ and decompose the equations in \eqref{Z22-L1} into the ones for the components. 

Noting the identities
\begin{equation}
	\partial_u = 2u \partial_y, \qquad \partial_{\bar u} = 2\bar{u} \partial_{\bar y}, \qquad \partial_{u\bar{u}} = 4 u\bar{u} \partial_{y\bar{y}},
\end{equation}
the LHS of the equations in \eqref{Z22-L1} yield
\begin{align}
	\partial_{u\bar u} \varphi_{00}
	&=4u\bar u \partial _{y\bar y} a_{00} 
	+2\bar u (\partial_{\bar y}+2y\partial_{y\bar y} )a_{10}
	+2u (\partial_{y}+2\bar y\partial_{y\bar y})b_{10}
	\notag\\
	&+(1+ 2y\partial_y +2\bar y\partial_{\bar y} +4y\bar y\partial_{y\bar y})b_{00},
	\notag\\[4pt]
	\partial_{u\bar u} \varphi_{11}
	&=4u\bar u \partial _{y\bar y} a_{11} 
	+  2\bar u (\partial_{\bar y}+2y\partial_{y\bar y} )a_{01}
	+2u (\partial_{y}+2\bar y\partial_{y\bar y})b_{01}
	\notag\\
	&+(1+ 2y\partial_y +2\bar y\partial_{\bar y} +4y\bar y\partial_{y\bar y})b_{11}.
	\label{Z22Liou-LHS}
\end{align}
These equations are simplified by introducing the [00]-graded commuting variables
\begin{equation}
	x := \sqrt{y}, \qquad \bar x := \sqrt{\bar y}. \label{DEF:xbarx}
\end{equation}
In terms of these variables \eqref{Z22Liou-LHS} is a power series of $  u x^{-1} $ and $\bar{u} \bar{x}^{-1}$:
\begin{align}
	\partial_{u\bar{u}}\varphi_{00} &= \partial_{x\bar x} (x \bar{x} b_{00}) + \frac{u}{x} \partial_{x\bar x} (\bar{x} b_{10}) + \frac{\bar u}{\bar x} \partial_{x\bar x} (x a_{10} )+ \frac{u \bar u}{x \bar x} \partial_{x\bar x} a_{00},
	\notag \\[4pt]
	\partial_{u\bar{u}}\varphi_{11} &= \partial_{x\bar x} (x \bar{x} b_{11}) + \frac{u}{x} \partial_{x\bar x} (\bar{x} b_{01}) + \frac{\bar u}{\bar x} \partial_{x\bar x} (x a_{01} )+ \frac{u \bar u}{x \bar x} \partial_{x\bar x} a_{11}. \label{Z22Liou-LHS2}
\end{align}
The RHS of \eqref{uubarcomp} are also expanded in the power series of $  u x^{-1} $ and $\bar{u} \bar{x}^{-1}$. From these, one obtains a system of eight equations for the component fields. 
Before presenting the system of equations, we take  the linear combination of the component fields
\begin{align}
	f_{00}^{\pm} &:= a_{00} \pm x \bar{x} b_{00}, 
	\qquad 
	f_{10}^{\pm} := x a_{10} \pm \bar{x} b_{10},
	\notag \\
	f_{11}^{\pm} &:= a_{11} \pm x \bar{x} b_{11}, 
	\qquad 
	f_{01}^{\pm} := x a_{01} \pm \bar{x} b_{01}. \label{DEF:Variablef}
\end{align}
Then the system of equations is presented in  the following form
	\begin{align}
		\partial_{x\bar x} f_{00}^{\pm} &= \pm e^{2f_{00}^{\pm}} 
		\big[ \cosh 2f_{10}^{\pm} \cosh 2f_{11}^{\pm} \cos 2f_{01}^{\pm} - \sinh 2f_{10}^{\pm} \sinh 2f_{11}^{\pm} \sin 2f_{01}^{\pm} \big],
		\notag \\[4pt]
		\partial_{x\bar x} f_{10}^{\pm} &= \pm e^{2f_{00}^{\pm}} 
		\big[ \sinh 2f_{10}^{\pm} \cosh 2f_{11}^{\pm} \cos 2f_{01}^{\pm} - \cosh 2f_{10}^{\pm} \sinh 2f_{11}^{\pm} \sin 2f_{01}^{\pm} \big],
        \notag \\[4pt]
		\partial_{x \bar x} f_{01}^{\pm} &= \pm e^{2f_{00}^{\pm}} 
		\big[ \sinh 2f_{10}^{\pm} \sinh 2f_{11}^{\pm} \cos 2f_{01}^{\pm} + \cosh 2f_{10}^{\pm} \cosh 2f_{11}^{\pm} \sin 2f_{01}^{\pm}\big],
		\notag \\[4pt]
		\partial_{x \bar x} f_{11}^{\pm} &= \pm e^{2f_{00}^{\pm}} 
		\big[ \cosh 2f_{10}^{\pm} \sinh 2f_{11}^{\pm} \cos 2f_{01}^{\pm} + \sinh 2f_{10}^{\pm} \cosh 2f_{11}^{\pm} \sin 2f_{01}^{\pm}\big].
		\label{LiouvillComp2}
	\end{align}
These are the equations for $\Z2$-graded functions whose arguments are [00]-graded commuting coordinates. \par
It should be pointed out that the same set of equations is derived from the alternative Lax pair defined by $L_v, L_{\bar{v}} $ . This is a consequence of the compatibility of the two conjugate sets of \eqref{laxes} Lax pairs which respect the $\Z2$-grading. \\
The \eqref{LiouvillComp2} system of equations is reduced to the Liouville equation if all the functions with non-trivial gradings are eliminated. Setting $ \varphi_{11} = 0 $ is equivalent to set $f_{11}^{\pm} = f_{01}^{\pm} = 0$. Then  \eqref{LiouvillComp2} is reduced to the following equations:
\begin{align}
		\partial_{x\bar{x}} f_{00}^{\pm} = \pm e^{2f_{00}^{\pm}} \cosh 2f_{10}^{\pm}, 
		\qquad
		\partial_{x\bar{x}} f_{10}^{\pm} = \pm e^{2f_{00}^{\pm}} \sinh 2f_{10}^{\pm}.
\end{align}
The Liouville equation is recovered by further setting $ f_{10} = 0$. Therefore, \eqref{LiouvillComp2} is a $\Z2$-graded extension of the Liouville equation which is, by construction, integrable.

\subsection{$\Z2$-Liouville equation in Matrix presentation} \label{SEC:MatLiouville}

The equation \eqref{Z22-L1} admits another interpretation if we use the matrix presentation. 

Let us introduce the matrix presentation of the coordinates, cf. \eqref{uMat}, \eqref{DuMat}
\begin{equation}
	u = M_1 z, \qquad \bar{u} = M_1 \bar{z}.
\end{equation}
It follows that
\begin{equation}
	\partial_u = M_1 \partial_z, \qquad \partial_{\bar{u}} = M_1 \partial_{\bar{z}} 
	\qquad
	\partial_{u\bar{u}} = \partial_{z\bar{z}} \mathbb{I}_4.
\end{equation}
This gives the matrix presentation of the component functions of  $\varphi_{00}(u,\bar{u})$ in \eqref{uubarcomp}, cf. \eqref{cIMat}
\begin{alignat}{2}
	a_{00}(y, \bar{y}) &= a(z^2,\bar{z}^2) \mathbb{I}_4,  &\qquad 
	a_{10}(y, \bar{y}) &= \alpha(z^2,\bar{z}^2) M_1, 
	\notag \\
	b_{00}(y, \bar{y}) &= b(z^2,\bar{z}^2) \mathbb{I}_4,  &\qquad 
	b_{10}(y, \bar{y}) &= \beta(z^2,\bar{z}^2) M_1 ,
\end{alignat}
where $ a, \alpha, b, \beta $ are non-graded complex functions. 
Therefore, one gets
\begin{equation}
	\varphi_{00}(u, \bar{u}) = \big[ a(z^2,\bar{z}^2) + z \alpha(z^2,\bar{z}^2) + \bar{z} \beta(z^2,\bar{z}^2) + z\bar{z} b(z^2,\bar{z}^2)  \big] \mathbb{I}_4 \equiv \varphi(z, \bar{z}) \mathbb{I}_4. \label{MatrixFields2}
\end{equation}
Similarly, one may write
\begin{align}
	 \varphi_{11}(u,\bar{u}) = \tilde{\varphi}(z,\bar{z}) M_3. \label{MatrixFields}
\end{align}

Inserting these expressions in \eqref{Z22-L1} we obtain the system of equations without $\Z2$-grading
\begin{equation}
	\partial_{z\bar{z}} \varphi = e^{2\varphi} \cosh 2\tilde{\varphi}, 
	\qquad
	\partial_{z\bar{z}} \tilde{\varphi} = e^{2\varphi} \sinh 2\tilde{\varphi}. \label{Z22Liouvill2}
\end{equation}
This system is equivalent to two decoupled Liouville equations,  as  seen from the positions:
\begin{equation}\label{decoupled}
	\phi_{\pm} := \varphi \pm \tilde{\varphi} 
	\quad \Rightarrow \quad 
	\partial_{z\bar{z}} \phi_{\pm} = e^{2\phi_{\pm}}.
\end{equation}
The system \eqref{Z22Liouvill2} is also equivalent to the split-complex $(\tilde{\mathbb{C}})$ version of the Liouville equation. Introducing the split-complex field 
\begin{align}
	\varphi_{\tilde{\mathbb{C}}}:= \varphi + j \tilde{\varphi}, \qquad j^2 = 1, \label{DEF:sComplexPhi}
\end{align}
then the two equations in \eqref{Z22Liouvill2} are combined into a single one:
	\begin{align}
		\partial_{z\bar{z}} \varphi_{\tilde{\mathbb{C}}} = \exp(2\varphi_{\tilde{\mathbb{C}}}).
	\end{align}

This result requires some comments. Even if, in the matrix presentation, one can obtain two decoupled Liouville equations from the fields $\varphi$, ${\tilde{\varphi}}$ which do not possess $\Z2$-grading, the original
 \eqref{LiouvillComp2} system of equations is non-trivial.  It consists of eight coupled $\Z2$-graded functions
$f_{ij}^\pm$ (for $i,j=0,1$) which cannot be linearly combined, as in \eqref{decoupled}, without breaking the $\Z2$ grading. The [$ij$] grading of the $f_{ij}^\pm$ functions plays an important physical role.  The [00] sector corresponds to ordinary bosons, while the [10], [01], [11] sectors correspond to parabosons which obey a different type of statistics, see \cite{top2}. It is the explicit expression of the  \eqref{LiouvillComp2} coupled system of equations which guarantees the compatibility of the derived $\Z2$-Liouville  equation with the $\Z2$-graded parastatistics.

%%%%%%%%%%%%%%%%%%%%%%%%%%%%%%%%%%%%%%%%%%%%%%%%%%%%%%%%%%%%%%%%%%%%%%%%%%%%%%%%%%%%%%%%%%%%%%%%
%
\setcounter{equation}{0}
\section{Zero-curvature formulation of $\Z2$-graded Sinh-Gordon model}

We construct the Toda system associated with the $\Z2$-graded affine algebra \affine introduced in \S \ref{SEC:affine}. 

Let us define the operator
\begin{equation}
	G := \frac{1}{2}H_0 + 2d_{00}.
\end{equation}
The elements of \affine have eigenvalues $ 0 $ or $ \pm 1 $  with respect to $ \mathrm{ad} G $, according to:
\begin{equation}
	\begin{array}{r|cccc}
		&\ [00]\ &\ [10]\ & [01] &\ [11] \ 
		\\ \hline
		+1 & & E_0^+, \ E_1^- & D_0^+, \ D_1^- & 
		\\[5pt]
		0 & H_0, \ d_{00}, \ c_{00} & & & Z_0, \ d_{11}, \ c_{11}
		\\[5pt]
		-1 & & E_0^-, \ E_{-1}^+, & D_0^-, \ D_{-1}^+
	\end{array}
\end{equation}
We introduce, following \cite{babo}, the [00]-graded fields
\begin{align}\label{gradedfields}
	\Phi(u,\bar{u}) &= \frac{1}{2} \varphi_{00} H_0 + \xi_{00}\, d_{00} + \frac{1}{2} \eta_{00}\, c_{00} +\frac{1}{2} \varphi_{11} Z_0 + \xi_{11} d_{11} + \frac{1}{2} \eta_{11} c_{11},
	\nonumber \\
	\Psi(v,\bar{v}) &= \frac{1}{2} \psi_{00} H_0 + \zeta_{00}\, d_{00} + \frac{1}{2} \rho_{00}\, c_{00} + \frac{1}{2} \psi_{11} Z_0 + \zeta_{11} d_{11} + \frac{1}{2} \rho_{11} c_{11}
\end{align}
and define 
	\begin{alignat}{2}
	L_u &=  -\partial_u \Phi + e^{\mathrm{ad} \Phi} \mathscr{E}_{+},
	& \qquad
	L_{\bar{u}} &= \partial_{\bar{u}} \Phi + e^{-\mathrm{ad} \Phi} \mathscr{E}_{-},
	\notag \\
	L_v &= - \partial_v \Psi  + e^{\mathrm{ad} \Psi} \mathscr{D}_{+},
	&
	L_{\bar{v}} &=  \partial_{\bar{v}} \Psi  + e^{-\mathrm{ad} \Psi} \mathscr{D}_{-},
\end{alignat}
where
\begin{equation}
	\mathscr{E}_{\pm} := E_0^{\pm} + E_{\pm 1}^{\mp}, \qquad 
	\mathscr{D}_{\pm} := D_0^{\pm} + D_{\pm 1}^{\mp}.
\end{equation}

The zero-curvature condition for $L_u, L_{\bar u}$, given by
\begin{equation}
	\partial_{\bar{u}} L_u - \partial_u L_{\bar{u}} + [ L_u, L_{\bar{u}}] = 0,  \label{ZeroC-sl2-2}
\end{equation}
is equivalent to
\begin{align}
	2 \partial_{u\bar{u}} \Phi &=[e^{\mathrm{ad} \Phi} \mathscr{E}_{+},e^{-\mathrm{ad} \Phi} \mathscr{E}_{-}] 
	\notag \\
	&= \big( e^{2\varphi_{00}} \cosh2\varphi_{11} - e^{2\xi_{00}-2\varphi_{00}} \cosh(2\varphi_{11}-2\xi_{11}) \big) H_0
	\nonumber \\
	&+ e^{2\xi_{00}-2\varphi_{00}} \cosh(2\varphi_{11}-2\xi_{11} ) \cdot c_{00}
	\nonumber \\
	&+\big( e^{2\varphi_{00}} \sinh2\varphi_{11} + e^{2\xi_{00}-2\varphi_{00}} \sinh(2\varphi_{11}-2\xi_{11}) \big) Z_0
	\nonumber\\
	&-e^{2\xi_{00}-2\varphi_{00}} \sinh(2\varphi_{11}-2\xi_{11}) \cdot c_{11}.
\end{align}
In terms of the graded fields on the RHS of (\ref{gradedfields}) one obtains:
\begin{align}
	\partial_{u\bar{u}} \varphi_{00} &= e^{2\varphi_{00}} \cosh2\varphi_{11} - e^{2\xi_{00}-2\varphi_{00}} \cosh(2\varphi_{11}-2\xi_{11}),
	\nonumber \\
	\partial_{u\bar{u}} \varphi_{11} &=e^{2\varphi_{00}} \sinh2\varphi_{11} + e^{2\xi_{00}-2\varphi_{00}} \sinh(2\varphi_{11}-2\xi_{11}),
	\nonumber\\
	\partial_{u\bar{u}} \eta_{00} &=e^{2\xi_{00}-2\varphi_{00}} \cosh(2\varphi_{11}-2\xi_{11} ),
	\nonumber \\
	\partial_{u\bar{u}} \eta_{11} &= -e^{2\xi_{00}-2\varphi_{00}} \sinh(2\varphi_{11}-2\xi_{11}),
	\nonumber \\
	\partial_{u\bar{u}} \xi_{00} &= \partial_{u\bar{u}} \xi_{11} = 0.
	\label{Affine-eq1}
\end{align}
Similar equations are obtained from the zero-curvature condition for $ L_v, L_{\bar v} $:
\begin{align}
	\partial_{v\bar{v}} \psi_{00} &=  e^{2\psi_{00}} \cosh2\psi_{11} - e^{2\zeta_{00}-2\psi_{00}} \cosh(2\psi_{11}-2\zeta_{11}),
	\nonumber \\
	\partial_{v\bar{v}} \psi_{11} &=  e^{2\psi_{00}} \sinh2\psi_{11} + e^{2\zeta_{00}-2\psi_{00}} \sinh(2\psi_{11}-2\zeta_{11}),
	\nonumber \\
	\partial_{v\bar{v}}  \rho_{00} &= e^{2\zeta_{00}-2\psi_{00}} \cosh(2\psi_{11}-2\zeta_{11}),
	\nonumber \\
	\partial_{v\bar{v}}  \rho_{11} &= -e^{2\zeta_{00}-2\psi_{00}} \sinh(2\psi_{11}-2\zeta_{11}),
	\nonumber \\
	\partial_{v\bar{v}} \zeta_{00} &= \partial_{v\bar{v}} \zeta_{11} = 0.
	\label{Affine-eq2}
\end{align}

Obviously, one may set $ \xi_{00} = \xi_{11} = \zeta_{00} = \zeta_{11} = 0$. 
Then,  \eqref{Affine-eq1} is reduced to
\begin{align}
	\partial_{u\bar{u}} \varphi_{00} &= 2 \sinh 2\varphi_{00}  \cosh 2\varphi_{11},
	\nonumber \\
	\partial_{u\bar{u}} \varphi_{11} &= 2 \cosh 2\varphi_{00} \sinh 2\varphi_{11},
	\nonumber\\
	\partial_{u\bar{u}} \eta_{00} &=e^{-2\varphi_{00}} \cosh 2\varphi_{11},
	\nonumber \\
	\partial_{u\bar{u}} \eta_{11} &= -e^{-2\varphi_{00}} \sinh 2\varphi_{11}.
	\label{Affine-eq3}
\end{align}
A similar reduction can be applied to \eqref{Affine-eq2}. \par
Since the dynamics of the fields $\eta_{00}, \eta_{11} $ is governed by $\varphi_{00}, \varphi_{11}$, we can focus only on the first two equations. 
They contain only fields of [00] and [11]-grading. 
However, by expanding $\varphi_{00}$ and $\varphi_{11}$ into component fields as we did in \S \ref{SEC:L-comp-eq}, a full set of $\Z2$-graded equations can be obtained. 

We employ the expansion \eqref{uubarcomp} and introduce the non-graded variables $x, \bar{x}$ defined in \eqref{DEF:xbarx}. After lengthy calculations, one obtains the following equations for the fields $ f^{\pm}_{\bm{a}}$ defined in  \eqref{DEF:Variablef}. 
From \eqref{Affine-eq1}
\begin{align}
		\partial_{x \bar x} f_{00}^{\pm} = & \pm 2 \sinh 2f_{00}^{\pm} \cosh 2f_{10}^{\pm} \cosh 2f_{11}^{\pm} \cos 2f_{01}^{\pm}
		\notag \\
		& \mp 2 \cosh 2f_{00}^{\pm} \sinh 2f_{10}^{\pm} \sinh 2f_{11}^{\pm} \sin 2f_{01}^{\pm},
		\notag \\[4pt]
		\partial_{x \bar x} f_{10}^{\pm} = & \pm 2 \cosh 2f_{00}^{\pm} \sinh 2f_{10}^{\pm}\cosh 2f_{11}^{\pm} \cos 2f_{01}^{\pm}
		\notag \\
		& \mp 2 \sinh f_{00}^{\pm} \cosh f_{10}^{\pm} \sinh 2f_{11}^{\pm} \sin 2f_{01}^{\pm} 
		\label{Affine-eq4}
\end{align}
and from \eqref{Affine-eq2}	
\begin{align}
		\partial_{x \bar x} f_{11}^{\pm} =& \pm 2 \cosh 2f_{00}^{\pm} \cosh 2f_{10}^{\pm} \sinh 2f_{11}^{\pm} \cos 2f_{01}^{\pm}
		\notag \\
		& \pm 2 \sinh 2f_{00}^{\pm} \sinh 2f_{01}^{\pm} \cosh 2f_{11}^{\pm} \sin 2f_{01}^{\pm},
		\notag \\[4pt]
		\partial_{x \bar x} f_{01}^{\pm} =& \pm 2 \sinh 2f_{00}^{\pm} \sinh 2f_{10}^{\pm} \sinh 2f_{11}^{\pm} \cos 2f_{01}^{\pm}
		\notag \\
		& \pm 2 \cosh 2f_{00}^{\pm} \cosh 2f_{10}^{\pm}\cosh 2f_{11}^{\pm} \sin 2f_{01}^{\pm}. \label{newaffine}
\end{align}
Setting  in \eqref{Affine-eq4} $\varphi_{11} = 0 $,  which is equivalent to $ f_{11}^{\pm} = f_{01}^{\pm} = 0 $, we get 
	\begin{align}
		\partial_{x\bar{x}} f_{00}^{\pm} &=  \pm 2 \sinh 2f_{00}^{\pm} \cosh 2f_{10}^{\pm},
		\\[4pt]
		\partial_{x\bar{x}} f_{10}^{\pm} &= \pm 2 \cosh 2f_{00}^{\pm} \sinh 2f_{10}^{\pm}.  
	\end{align}
Under the further position $ f^{\pm}_{10} = 0$ we recover the Sinh-Gordon equation. 

We can also  consider the matrix presentation of the system \eqref{Affine-eq3}. 
As done in \S \ref{SEC:MatLiouville}, the fields $\varphi_{00}, \varphi_{11} $ can be written as a product of non-graded functions coupled with matrices, see \eqref{MatrixFields2}, \eqref{MatrixFields}. 
Then, the first two equations in \eqref{Affine-eq3} yield
\begin{equation}
	\partial_{z\bar{z}} \varphi = 2\sinh 2\varphi \cosh 2\tilde{\varphi}, 
	\qquad
	\partial_{z\bar{z}} \tilde{\varphi} = 2\cosh 2\varphi \sinh 2\tilde{\varphi}.
	\label{SGMatrixForm}
\end{equation}
These equations are equivalent to two decoupled Sinh-Gordon equations:
\begin{equation}
	\phi_{\pm} := \varphi \pm \tilde{\varphi} 
	\quad \Rightarrow \quad 
	\partial_{z\bar{z}} \phi_{\pm} = 2 \sinh 2\phi_{\pm}.
\end{equation}
The equations \eqref{SGMatrixForm} are also equivalent to
\begin{equation}
	\partial_{z \bar z} \varphi_{\tilde{\mathbb{C}}} = 2\sinh \varphi_{\tilde{\mathbb{C}}},
\end{equation}
where $\varphi_{\tilde{\mathbb{C}}}$ is defined in \eqref{DEF:sComplexPhi} in terms of a split-complex number. 
\\
{\color{\BL}{The \eqref{Affine-eq4}, \eqref{newaffine} system of coupled equations for eight $\Z2$-graded 
functions, obtained from \affine after setting $ \xi_{00} = \xi_{11} = \zeta_{00} = \zeta_{11} = 0$, is the integrable
$\Z2$-graded extension of the Sinh-Gordon equation. The interpretation of the results  parallels what already discussed in the construction of the $\Z2$-Liouville model.
 }}

%%%%%%%%%%%%%%%%%%%%%%%%%%%%%%%%%%%%%%%%%%%%%%%%%%%%%%%%%%%%%%%%%%
\section{Conclusions}
\setcounter{equation}{0}

The paper presented integrable $\Z2$-graded extensions of both classical Liouville and Sinh-Gordon equations, obtained as systems of eight coupled $\Z2$-graded functions obeying a color Lie algebra parastatistics. Besides the obtained results, a general framework was presented to covariantly define Lax pair formulations for $\Z2$-graded extensions of finite semisimple and affine Lie algebras. 
In this paper we solved subtle issues like the introduction of graded coordinates and the proper formulation of the theory in terms of conjugate Lax pairs which produce compatible equations. 
Even if here we only explicitly worked the $\Z2$-graded extension of $sl_2$, these features can be easily applied to derive Toda field models induced by generic $\Z2$-graded Lie algebras; the most general construction is left for a future work.  
%Subtle issues had to be solved like the introduction of graded coordinates, the formulation in terms of conjugated Lax pairs which produce compatible equations and so on. 
Following the original \cite{lezsav} construction, the reconstruction theorem in \cite{baboto} and the \cite{babo} extension to affine Lie algebras, the introduction of the covariant Lax pair formulation guarantees the integrability of the models.  We also introduced the alternative derivation of the Liouville model in terms of a $\Z2$-graded version of the \cite{polyak} Polyakov's soldering procedure.  \par
For the Liouville extension, the covariant Lax pair formulation was based on the finite
$\Z2$-graded color Lie algebra \algnospace ; for the Sinh-Gordon extension, on the affine $\Z2$-graded color Lie algebra \affine which admits two central charges, one of them [$11$]-graded.\par
 The \affine algebra plays a role in the $\Z2$-Liouville theory as well; it generates the transformations of the WZNW currents which, under Hamiltonian reduction, produce the $\Z2$-Liouville equation. It is interesting to note that, in this application, only one central charge appears (the [$11$]-graded central charge is vanishing).  By imposing the Hamiltonian reduction, the current algebra induces a $\Z2$-graded version of the Virasoro algebra. \\
The classical theories under investigation can be easily expressed in a $\Z2$-graded Lagrangian formulation following \cite{akt1} and quantized with the prescriptions discussed in \cite{akt2}.}

Further lines of research consist in extending the zero-curvature formulation to $\Z2$-graded superToda theories
derived from $\Z2$-graded color Lie {\textit{super}}algebras; the simplest of such type of models is obtained from
the $\Z2$-graded $osp(1|2)$ superalgebra whose irreps are studied in \cite{amai}. The extra ingredient to take into account, with respect to the $\Z2$-graded color Lie algebra formulation, is the introduction of
(para)Grassmann coordinates.\par
It should be finally mentioned that, quite likely, the $\Z2$-graded Lax pair formulation could be adapted  to introduce $\Z2$-graded non-abelian Toda field theories, mimicking the construction presented in \cite{gesa}.  For these models the dynamical fields are no longer associated with the Cartan sector of a semisimple Lie algebra $g$, but with a non-abelian subalgebra.
~\\

 {\Large{\bf Acknowledgments}}
{}~\par{}~\\
Z. K. and F. T. are grateful to the Osaka Metropolitan University, where this work has been completed, for the hospitality. \\ N. A. is supported by JSPS KAKENHI Grant Number JP23K03217. \\
F. T. is supported by CNPq (PQ grant 308846/2021-4).

\end{document}